\documentclass[prl,twocolumn,showpacs,preprintnumbers,amsmath,amssymb]{revtex4}

\usepackage{graphicx}
\usepackage{dcolumn}
\usepackage{bm}

\begin{document}

\preprint{APS/123-QED}

\title{Ab-initio description of the magnetic shape anisotropy 
       due to the Breit interaction}

\author{S.~Bornemann, J.~Min\'ar, J.~Braun, D.~K\"odderitzsch and  H.~Ebert}

\affiliation{Department Chemie und Biochemie, 
             Ludwig-Maximilians-Universit\"at M\"unchen, 
             81377 M\"unchen, Germany}

\date{\today}

\begin{abstract}
A quantum-mechanical description of the magnetic shape anisotropy, that is
usually ascribed to the classical magnetic dipole-dipole interaction, has been
developed. This is achieved by including the Breit-interaction, that can be
seen as an electronic current-current interaction in addition to the
conventional Coulomb interaction, within fully relativistic band structure
calculations. The major sources of the magnetic anisotropy, spin-orbit coupling
and the Breit-interaction, are treated coherently this way. This seems to be
especially important for layered systems for which often both sources
contribute with opposite sign to the magnetic anisotropy energy. Applications
to layered transition metal systems are presented to demonstrate the
implications of this new approach in treating the magnetic shape anisotropy.
\end{abstract}

\maketitle

Magnetic anisotropy is among the most important properties of magnetic
materials in particular concerning their application in devices. When
discussing the magnetic anisotropy energy of a material, denoting the
difference in energy for two orientations of the magnetisation, an incoherent
approach is used so far~\cite{Bru93,Blu99}. On the one hand side, the
dependency of the electronic structure and the associated total energy on the
orientation of the magnetisation, that is induced by spin-orbit coupling (SOC),
is accounted for by corresponding relativistic band structure calculations. On
the other hand, the additional shape anisotropy is ascribed to the anisotropy
of the magnetic dipole-dipole coupling, that is treated in a classical way.
This hybrid approach is used in particular when dealing with layered transition
metal systems. The pioneering and successful theoretical work on magnetic
surface films by Gay and Richter~\cite{GR86} was followed later on by many more
investigations that benefited from the extension of standard band structure
schemes to account simultaneously for the presence of spin-orbit coupling and
spin-polarisation, i.e.\ magnetisation, in a numerically reliable way.
Especially interesting in this context are investigations on systems showing a
competition of the SOC and shape induced contributions to the magnetic
anisotropy energy.  This situation is frequently encountered for magnetic
multi-layer or surface layer systems for which a flip of the magnetic easy axis
from out-of-plane to in-plane may be observed when the thickness of the
magnetic layer is increased starting from a single mono-layer.  In fact
corresponding experimental findings could be reproduced by calculations based
on the above mentioned hybrid scheme in the case of magnetic
multi-layers~\cite{DKS90} as well as surface layer systems~\cite{SUW95}. 

In spite of the successful applications of this hybrid scheme in dealing with
the magnetic anisotropy, one has to keep in mind that there is no real
justification for its use.  Furthermore, as there have been no coherent quantum
mechanical investigations performed so far, there is no experience on the range
of its applicability.  It seems that the first steps towards a coherent quantum
mechanical description of the magnetic anisotropy were made by
Jansen~\cite{Jan88,Jan88a}, who pointed out that the shape anisotropy is
ultimately caused by the Breit interaction -- a relativistic correction to the
Coulomb interaction between moving electrons~\cite{BS77,BS57}. While Jansen
performed model calculations to investigate the magnetic anisotropy energy
contribution caused by the Breit interaction, first numerical investigations
were done by Stiles et al.~\cite{SHHZ01}. However, these authors restricted to
the spin-other-orbit part of the Breit interaction that implies a
current-current interaction.  Performing a Gordon decomposition of the current
density into a spin and orbital part~\cite{Ros61} one gets three additional
terms. Moreover, Stiles et al. investigated the pure bulk ferromagnetic metals
bcc-Fe, hcp-Co and fcc-Ni for which the shape anisotropy is rather small. As a
consequence,  the results of this first numerical work are not very conclusive.

In this contribution we present a fully relativistic description of the
magnetic anisotropy that accounts for the shape anisotropy by incorporating the
full Breit interaction within the Dirac equation for magnetic solids. The
scheme has been implemented by using a fully relativistic multiple-scattering
or Korringa-Kohn-Rostoker (KKR) formalism.  To demonstrate the power of this
new approach applications to the layered systems Fe$_n$Pd$_n$ and
Fe$_n$/Au(001) are presented. The results for the magnetic anisotropy energy
are compared to results obtained using the conventional hybrid approach.

The Breit-interaction \cite{BS77,BS57} is a correction to the Coulomb interaction
between moving electrons that may be split into its magnetic and retardation
part:
\begin{eqnarray}
\mathcal H^{(1,2)}_{\mathrm {BI}} 
&=& -\frac{e^2}{r_{12}} \vec{\alpha}_1 \cdot \vec{\alpha}_2 \nonumber\\
& & +\frac{e^2}{2r_{12}} \Bigl[  \vec{\alpha}_1 \cdot \vec{\alpha}_2
- \left(\vec{\alpha}_1 \cdot \vec{e}_{12} \right)
\left(\vec{\alpha}_2 \cdot \vec{e}_{12}\right) \Bigr] \; .
\label{H_Breit}
\end{eqnarray}
Within this relativistic formulation, the standard Dirac matrices $\alpha_i$
are  connected to the current density operator  $j_i$ via $\vec
j=-|e|c\vec{\alpha}$~\cite{Ros61} and $r_{12}$ denotes the distance between two
electrons.  The Gordon decomposition of the electronic current mentioned above
is not used in the following. This implies that all coupling mechanisms
connected with the Breit interaction are finally accounted for.

In the following we focus here on the magnetic part as the retardation term
does not contribute on the Hartree level~\cite{MJ71}.  This allows to represent
the interaction of an electron with all the other electrons by the vector
potential:
\begin{eqnarray}
\vec{A}(\vec{r}\,) = \frac{1}{c}\int d^3r' \frac{\vec{j}(\vec{r}\,')}{|\vec{r}-\vec{r}\,'|} \; .
\label{currentint}
\end{eqnarray}
Including the corresponding interaction term within electronic structure
calculations done in the framework of relativistic spin density functional
theory (SDFT) \cite{MV79} leads to the following single particle Dirac Hamiltonian:
\begin{equation}
\mathcal H_{D}=-ic\vec{\alpha}\cdot\vec{\nabla}  
 +\beta mc^2 +V_{\rm eff}(\vec{r}\,) + \beta\vec{\Sigma}\cdot \vec{B}(\vec{r}\,) 
 + e \vec{\alpha} \cdot \vec{A}(\vec{r}\,)  \;, \label{Hdirac}
\end{equation}
with the corresponding total energy $E$ given by:
\begin{equation}
\label{Etot}
E  =  T_{\rm s} + E_{\rm ext} + E_{\rm H}^{\rm C} + E_{\rm H}^{\rm T} + E_{\rm xc}  
\end{equation}
Here the first three and the last terms have their usual meaning, i.e.\ they
stand for the kinetic energy ($T_{\rm s}$), the electron-nucleus  ($E_{\rm
ext}$) and electron-electron ($E_{\rm H}^{\rm C}$) Coulomb interaction and the
exchange-correlation ($E_{\rm xc}$) contributions.  The fourth term:
\begin{equation}
E_{\rm H}^{\rm T} = -\frac{e^2}{2c} \int d^3r \int d^3r' \,
\frac{\vec{j}(\vec{r}\,)\vec{j}(\vec{r}\,')}{|r-r'|} \; ,
\end{equation}
going beyond standard SDFT, is a Hartree-like term due to the Breit or
current-current interaction.  As mentioned above the retardation term in
Eq.~(\ref{H_Breit}) does not contribute to $E_{\rm H}^{\rm T}$~\cite{MJ71}.
This could indeed be veryfied by our numerical results.

As usually done for total energy calculations, the kinetic energy term in
Eq.~(\ref{Etot}) can be eliminated by making use of the Hamiltonian in
Eq.~(\ref{Hdirac}).

When calculating the vector potential $\vec{A}(\vec{r}\,)$ it is advantageous
to decompose the integration regime in Eq.~(\ref{currentint}) into the central
atomic cell $i$ and its surrounding giving rise to the on- and off-site
contribution, respectively, for the vector potential within cell $i$:
\begin{eqnarray}
\vec{A}_i(\vec{r}\,) &=& \vec{A}^{\mathrm{on}}_i(\vec{r}\,) 
                      +\vec{A}^{\mathrm{off}}_i(\vec{r}\,)     \; .
\end{eqnarray}
%
The  on-site contribution  $\vec{A}^{\mathrm{on}}_i$ can be determined directly
from the currents within atomic cell $i$, on the basis of Eq.~(\ref{H_Breit}).
The off-site contribution $\vec{A}^{\mathrm{off}}_i$ from all other sites can
be obtained by applying the common far field approximation
\begin{eqnarray}
\vec{A}^{\mathrm{off}}_i(\vec{r}\,) &=&
\sum_{j\ne i} \frac{\vec{M}_j\times(\vec{r}-\vec{R}_j)}{|\vec{r}-\vec{R}_j|^3} 
\; ,
\label{latticesum}
\end{eqnarray}
where the total magnetic moment $\vec{M}_j$ represents the current distribution
in atomic cell $j$. The lattice summation in Eq.~(\ref{latticesum}) is dealt
with by an Ewald summation technique in the case of two- or three-dimensional
periodic systems.

Using a band structure method based on a decomposition of the system into
atomic cells, as for the KKR-method used here, Eq.~(\ref{Hdirac}) has to be
solved in a first step for isolated atomic cells (single-site problem). For
this purpose it is helpful to expand the vector potential by means of spherical
harmonics. For the Breit-interaction term, the last term in Eq.~(\ref{Hdirac}),
one obtains:
\begin{equation}
\mathcal H_{\mathrm {BI}}   
      = e \sum_{m_\alpha} \alpha_{m_\alpha}  
          \sum_{\ell_A m_A} A^{-m_\alpha}_{\ell_A m_A}(r) 
          Y^{m_A}_{\ell_A}(\hat{r}) \;. \label{Breitvec}
\end{equation}

In the implementation presented here the atomic sphere approximation (ASA)
together with a restriction to collinear magnetism has been applied, i.e.\
within an atomic cell one has $V_{\rm eff}(\vec{r}\,)= V(r)$ and
$\vec{B}(\vec{r}\,)= B(r)\hat{e}_M$, with the orientation of the magnetic
moments $\hat{e}_M$ that fixes the local $z$-axis. In line with these
geometrical simplifications $\vec{A}(\vec{r}\,)$ is restricted to have
rotational symmetry around $\hat{e}_M$ with $\vec{A}(\vec{r}\,)$ pointing
everywhere in tangential direction, i.e.\  $\vec{A}(\vec{r}\,)=
A(r,\theta)\hat{e}_\phi$ implying $A_{\ell_A -1}^{+1}(r) = -A_{\ell_A
+1}^{-1}(r)$ and all other terms being zero. (Further technical details will be
given elsewhere \cite{BMBK+11}).

With the single-site Dirac equation  being solved, the electronic structure of
the investigated system can be calculated by means of the standard relativistic
multiple scattering or KKR-method \cite{Ebe00}.  Calculating the total energy
$E(\hat{n})$  on the basis of Eq.~(\ref{Etot}) for two different orientations
of the magnetisation $\hat{n}$ and $\hat{n}'$, respectively, gives the
corresponding magnetic anisotropy energy $\Delta E$ as the difference
$E(\hat{n}) - E(\hat{n}')$. 

\begin{figure}
\includegraphics[height=5cm,clip]{Fe_mono_vecpot_comp.eps}
\includegraphics[height=5cm,clip]{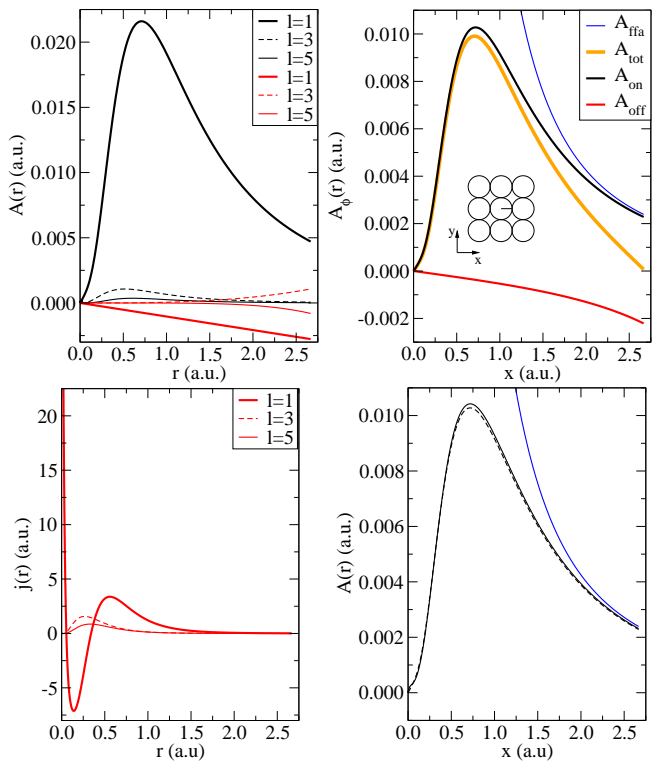}
\includegraphics[height=5cm,clip]{Fe_mono_currentdensity.eps}
\hspace{0.25cm}
\includegraphics[height=5cm,clip]{Fe_mono_vecpot.eps}
\caption{\label{fig:Fe_mono_vecpot1}
Vector potential and current density for Fe in a free-standing monolayer.  Top
left: On- (black) and off-site (red) contributions to the vector potential
functions $A^{-m_A}_{\ell_A m_A}(r)$ for $\ell_A=1,3,5$ and $m_A=1$ in
atomic units. Top right: On- (black) and off-site (red) $A_{\phi}$ component
along the $x$-direction ($\theta=\frac{\pi}{2}$,$\phi=0$).
$\vec{A}^{\mathrm{on}}$ is compared to the far field approximation
$\vec{A}^{\mathrm{ffa}}$ (blue curve).  Bottom left: radial current density
functions $j^{-m}_{\ell m}(r)$ for $\ell=1,3,5$ and $m=1$.  Bottom right:
comparison of $A_{\phi}$ along $x$ calculated via $\mathcal H_{\mathrm
{BI}}$ (full line) and from the current density (dashed line).
}
\end{figure}

As an   application of the approach sketched above we consider the
geometrically simple case of a free-standing bcc Fe monolayer with a lattice
constant of bulk Fe ($a= 2.87~\AA$).  The magnetic moment  $\vec{M}$ was taken
to point out-of-plane, i.e.\ along the $z$-axis of the system.  The top left
panel of Fig.~\ref{fig:Fe_mono_vecpot1}  shows the resulting radial vector
functions $A^{-m_A}_{\ell_A m_A}(r)$ for $m_A=1$ with the on- and off-site
parts to  $\vec{A}(\vec{r}\,)$ indicated by black and red lines, respectively,

The top right panel of Fig.~\ref{fig:Fe_mono_vecpot1} shows the resulting
radial dependence of the $\phi$-component $A_{\phi}^{\mathrm{on}}$ along the
$x$-direction (black line) together with $A_{\phi}^{\mathrm{off}}$ obtained via
Eq.~(\ref{latticesum}) (red line).  Applying the far field approximation of
Eq.~(\ref{latticesum}) also to the on-site term results in the blue curve which
is divergent at the origin. In the outermost region of an atomic sphere,
however, the far field approximation to $\vec{A}(\vec{r}\,)$ is already very
good justifying the use of Eq.~(\ref{latticesum}) also for nearest neighbouring
atoms.

In order to verify the correctness of $A_{\phi}^{\mathrm{on}}$ when calculated
directly via $\mathcal H_{\mathrm {BI}}$ (see Eq.~(\ref{H_Breit})) we also
determined the current density $\vec{j}(\vec{r}\,)$ within an Fe sphere and
then computed $A_{\phi}^{\mathrm{on}}$ via Eq.~(\ref{currentint}).  The bottom
left panel of Fig.~\ref{fig:Fe_mono_vecpot1} shows the radial electronic
current density distribution functions $j^{-m}_{\ell m}(r)$ (defined in analogy
to $A^{-m_A}_{\ell_A m_A}$ in Eq.~(\ref{Breitvec})) for an Fe atom and the
bottom right panel of Fig.~\ref{fig:Fe_mono_vecpot1} presents a comparison of
$A_{\phi}^{\mathrm{on}}$ resulting from $j^{-m}_{\ell m}(r)$ (dashed line) and
obtained from $\mathcal H_{\mathrm {BI}}$. As one can see both approaches give
nearly identical results.

\begin{figure}[htb]
\begin{center}
\includegraphics[width=0.95\columnwidth,clip]{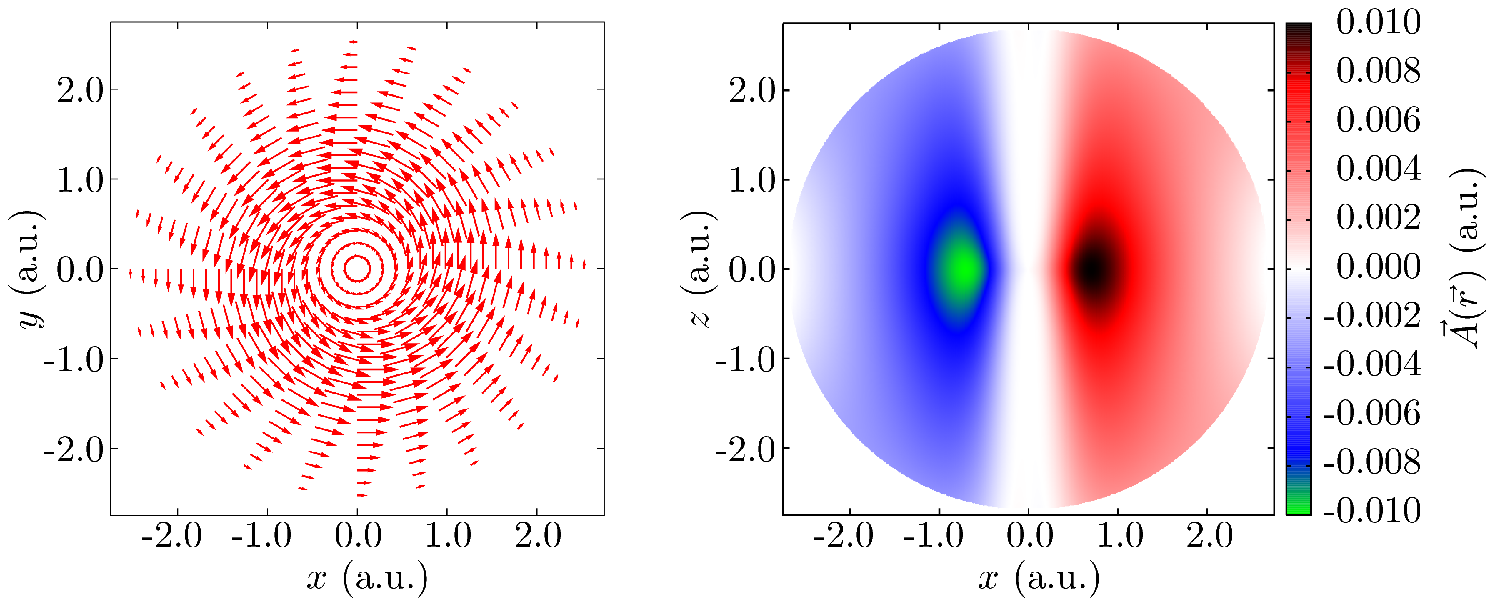}
\includegraphics[width=0.95\columnwidth,clip]{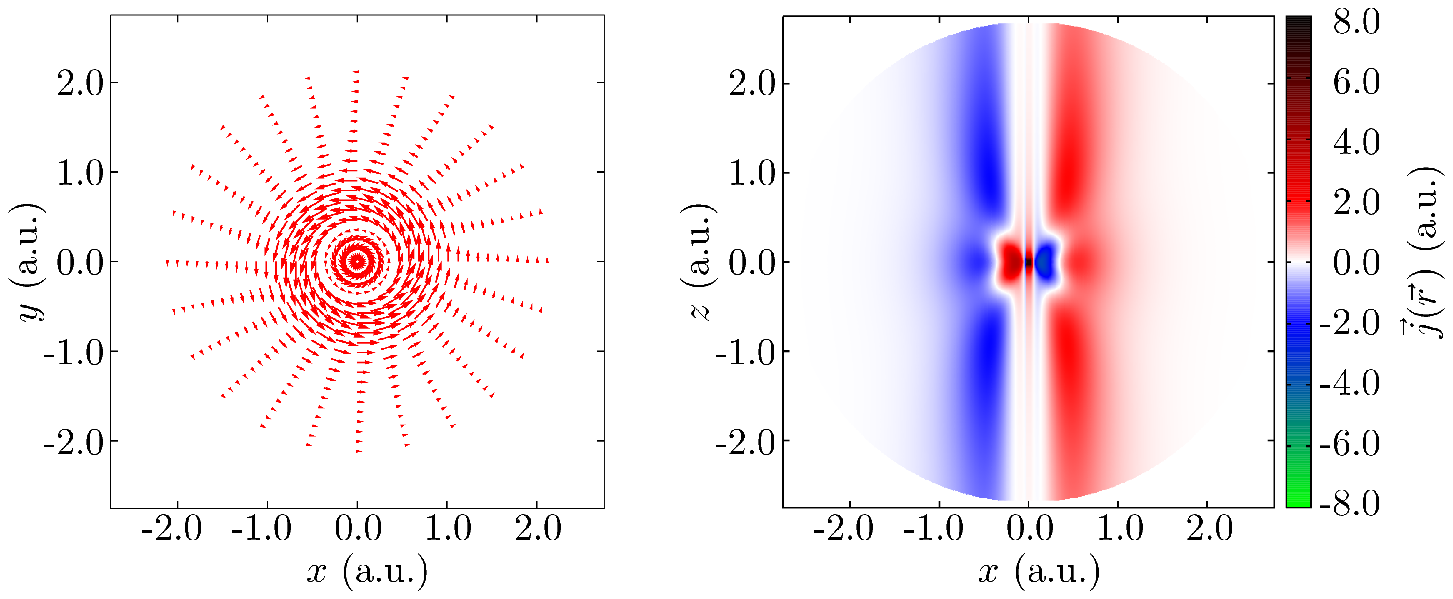}
\end{center}
\caption{\label{fig:Fe_mono_vecpot2}
Top: Vector potential $\vec A(\vec r)$ of Fe within the $xy$-plane with $z=0$ 
(left) and within the $xz$-plane with $y=0$ (right) for the 
magnetisation along the $z$-axis. Bottom: Corresponding current density $\vec j(\vec r)$
within the $xy$-plane (left) and within the $xz$-plane (right).
}
\end{figure}

A further impression of the spatial variation of $\vec{A}(\vec{r}\,)$ and
$\vec{j}(\vec{r}\,)$ is given in Fig.~\ref{fig:Fe_mono_vecpot2} showing the
vector fields in the $xy$-plane and their color-coded amplitude in the
$xz$-plane. The figure reflects the rotational symmetry of $\vec{A}(\vec{r}\,)$
imposed by the use of the ASA (atomic sphere approximation) and the alignment
of the magnetisation along the $z$-direction.
%

For the anisotropy energy $\Delta E = E(\hat{x}) - E(\hat{z})$ of the
free-standing bcc Fe monolayer we obtain $-0.063$~meV with the sign indicating
that the magnetisation favours an in-plane orientation.  $\Delta E$ can be
decomposed into $\Delta E_{\mathrm{SOC}}=0.096$~meV preferring an out-of-plane
magnetic easy axis and a dominating $\Delta E_{\mathrm{BI}}=-0.159$~meV causing
the magnetisation to lie in-plane. Here, $\Delta E_{\mathrm{BI}} = \Delta E -
\Delta E_{\mathrm{SOC}}$ has been obtained by performing calculations with
(SOC+BI) and without (SOC only) $\mathcal H_{\mathrm {BI}}$ giving $\Delta E$
and $\Delta E_{\mathrm{SOC}}$, respectively.  For the classical dipolar shape
anisotropy $\Delta E_{\mathrm{dd}}$ we get $-0.169$~meV which agrees
astonishingly well with the quantum mechanical $\Delta E_{\mathrm{BI}}$.

A second application dealt with the multilayer systems Fe$_n$Pd$_n$. For these
a fcc structure with (001)-oriented atomic layers has been assumed, i.e.\ with
the $z$-axis perpendicular to the Fe- and Pd-layers, respectively. For the
special case $n=1$ this corresponds to the CuAu-structure. 

\begin{figure}
\includegraphics[width=0.80\columnwidth,clip]{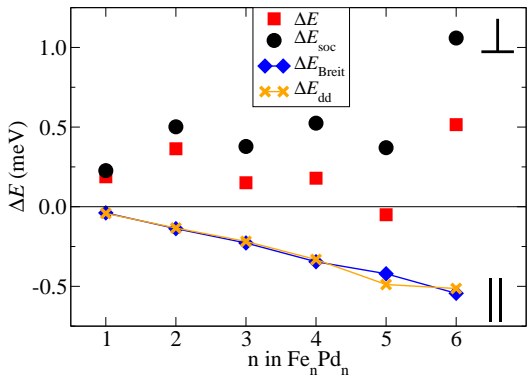}
\caption{\label{fig:FePd}Magnetic anisotropy energies for repeating
Fe$_n$Pd$_n$ multilayers as function of $n$: 
total magnetic anisotropy energy $\Delta E$
(squares) and its decomposition into magneto-crystalline part $\Delta E_{\rm SOC}$
(circles) and Breit part $\Delta E_{\rm BI}$ (diamonds) which is compared to its
classical approximation (crosses).}
\end{figure}

The impact of the Breit-interaction on the electronic structure leads to a
competition of the various contributions to the magnetic anisotropy energy
$\Delta E$ of Fe$_n$Pd$_n$ when the parameter $n$ is varied. Using the
conventional approach that accounts only for $\Delta E_{\rm SOC}$ within the
electronic structure calculations leads to a strong contribution that favors an
out-of-plane orientation of the magnetisation for $n=1-6$ (see
Fig.~\ref{fig:FePd}).  The additional classical shape anisotropy contribution
$\Delta E_{\rm dd}$ gives rise to a contribution that favours an in-plane
orientation of the magnetisation.  As one may expect, $\Delta E_{\rm dd}$ of
Fe$_n$Pd$_n$ is primarily determined by the magnetic moments within the Fe
layers that amounts to be between 2.74 and 2.95~$\mu_B$ for all values of $n$
considered here. The induced moments on the Pd layers that are in the range
0.005 to 0.330~$\mu_B$ are only of minor importance.  As a result of this
$\Delta E_{\rm dd}$ increases nearly linearly with $n$ for the range of $n$
considered here. For $n=5$ and presumably also for $n>6$ the magnetic
dipole-dipole term exceeds the SOC-induced term leading to a flip of the
magnetic easy axis from out-of-plane to in-plane. 

Calculating  the anisotropy energy $\Delta E$ on the basis of the coherent
SOC+BI scheme, $\Delta E$ follows qualitatively the variation of the
SOC-induced magnetic anisotropy energy with $n$ and implies also a flip of the
magnetic easy axis (see Fig.~\ref{fig:FePd}).  To allow for a direct comparison
of the two approaches the difference $\Delta E_{\rm BI}$ of SOC+BI and the
SOC-only scheme is shown as well in Fig.~\ref{fig:FePd}. As can be seen,
$\Delta E_{\rm BI}$ is very close to the classical $\Delta E_{\rm dd}$.  This
result obviously justifies the use of the conventional classical approach for
the shape anisotropy used so far. In particular the conventional scheme seems
to reproduce the quantum mechanical result not only in a qualitative but in
general also quantitatively in a satisfying way. Obviously, only for rather
short interatomic distances  one has to be aware of possibly pronounced
deviations between the classical and quantum mechanical approaches.
\begin{figure}
\includegraphics[width=0.80\columnwidth,clip]{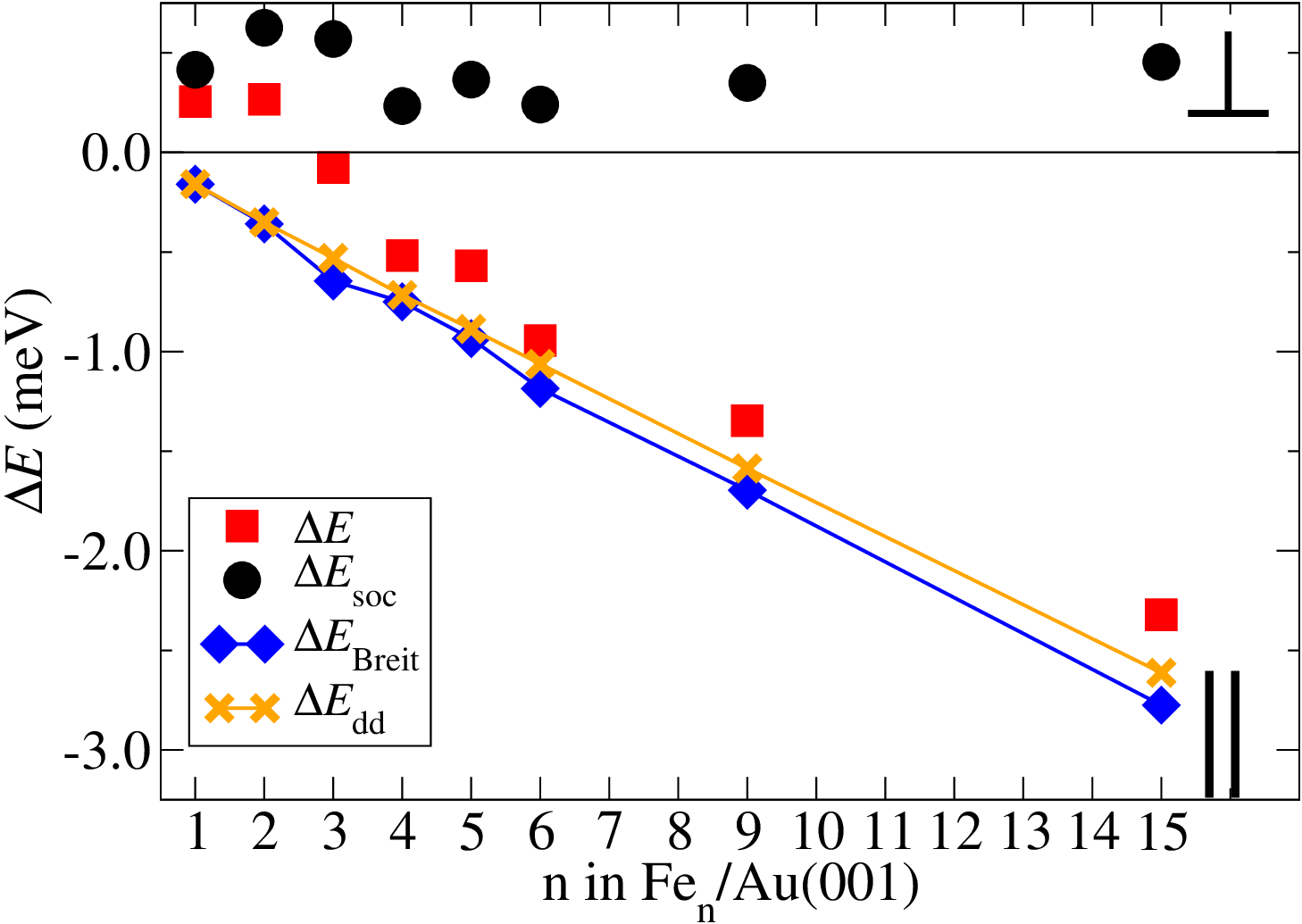}
\caption{\label{fig:FeAu001} As for Fig.~\ref{fig:FePd}, but for the 
surface layer sytem  Fe$_n$/Au(001).}
\end{figure}
These conclusions are confirmed by results obtained for the closely related
multilayer systems Fe$_n$Pt$_n$, Co$_n$Pd$_n$ and Co$_n$Pt$_n$ as well as first
applications to surface layer systems. Fig.~\ref{fig:FeAu001} shows
corresponding results for the system Fe$_n$/Au(001) that exhibits a flip of the
easy axis from out-of-plane to in-plane for around three monolayers of Fe (see
also Ref.~\cite{SUW95}). Again the contribution $\Delta E_{\rm BI}$ to the
total magnetic anisotropy energy  $\Delta E$ is found to be very close to the
classical result $\Delta E_{\rm dd}$. Only for rather large thicknesses a
noteworthy deviation of the two can be seen in Fig.~\ref{fig:FeAu001}.

In summary the Breit interaction has been incorporated within fully
relativistic  band structure calculations for magnetic layered systems. This
development gives access to an ab-initio calculation of the magnetic shape
anisotropy energy using a coherent approach that accounts simultaneously for
spin-orbit coupling and the Breit interaction. First applications of this new
approach to the systems Fe$_n$Pd$_n$ and Fe$_n$/Au(001) were presented. Taking
the difference of the calculated magnetic anisotropy energy obtained via the
combined SOC+BI - and the SOC-only approaches, the contribution due to the
Breit interaction could be separated.  For all systems investigated so far it
was found that the resulting Breit contribution is very close to the classical
result calculated on the basis of the magnetic dipole-dipole interaction. This
result can be explained to some extend by the fact that it is the magnetic part
of the Breit interaction that gives rise to the shape anisotropy while the
retardation term does not contribute on the Hartree-level.

Financial support by the DFG (Deutsche Forschungsgemeinschaft) project EB
154/14-2 is gratefully acknowledged.

\bibliographystyle{prsty}

\begin{thebibliography}{10}

\bibitem{Bru93}
P. Bruno,  in {\em Magnetismus von Festk\"orpern und Grenzfl\"achen}, edited by
  {\protect Forschungszentrum~J\"ulich~GmbH, Institut f\"ur
  Festk\"orperforschung} (Forschungszentrum J\"ulich GmbH, J\"ulich, 1993), p.\
  24.1.

\bibitem{Blu99}
S. Bl\"ugel,  in {\em 30.\ Ferienkurs des Instituts f\"ur Festk\"orperforschung
  1999 ''Magnetische Schichtsysteme''}, edited by {\protect
  Forschungszentrum~J\"ulich~GmbH, Institut f\"ur Festk\"orperforschung}
  (Forschungszentrum J\"ulich GmbH, J\"ulich, 1999), p.\ C1.1.

\bibitem{GR86}
J.~G. Gay and R. Richter, Phys. Rev. Lett. {\bf 56},  2728  (1986).

\bibitem{DKS90}
G.~H.~O. Daalderop, P.~J. Kelly, and M.~F.~H. Schuurmans, Phys. Rev. B {\bf
  41},  11919  (1990).

\bibitem{SUW95}
L. Szunyogh, B. \'Ujfalussy, and P. Weinberger, Phys. Rev. B {\bf 51},  9552
  (1995).

\bibitem{Jan88}
H.~J.~F. Jansen, J. Appl. Physics {\bf 64},  5604  (1988).

\bibitem{Jan88a}
H.~J.~F. Jansen, Phys. Rev. B {\bf 38},  8022  (1988).

\bibitem{BS77}
W.~M. Bibby and D. Shoenberg, Phys. Letters {\bf 60A},  235  (1977).

\bibitem{BS57}
H. Bethe and E. Salpeter, {\em Quantum Mechanics of One- and Two-Electron
  Atoms} (Springer, New York, 1957).

\bibitem{SHHZ01}
M.~D. Stiles, S.~V. Halilov, R.~A. Hyman, and A. Zangwill, Phys. Rev. B {\bf
  64},  104430  (2001).

\bibitem{Ros61}
M.~E. Rose, {\em Relativistic Electron Theory} (Wiley, New York, 1961).

\bibitem{MJ71}
J.~B. Mann and W.~R. Johnson, Phys. Rev. A {\bf 4},  41  (1971).

\bibitem{MV79}
A.~H. MacDonald and S.~H. Vosko, J. Phys. C: Solid State Phys. {\bf 12},  2977
  (1979).

\bibitem{BMBK+11}
S. Bornemann, J. Min\'ar, J. Braun, D. K\"odderitzsch, and H. Ebert
  (unpublished).

\bibitem{Ebe00}
H. Ebert,  in {\em Electronic Structure and Physical Properties of Solids},
  Vol.~535 of {\em Lecture Notes in Physics}, edited by H. Dreyss\'{e}
  (Springer, Berlin, 2000), p.\ 191.

\end{thebibliography}

\end{document}